\documentclass[twocolumn,english,aps,prl]{revtex4}
\usepackage[T1]{fontenc}
\usepackage[latin9]{inputenc}
\usepackage{xcolor}
\usepackage{pdfcolmk}
\usepackage{amsmath}
\usepackage{amssymb}
\usepackage{graphicx}
\usepackage{esint}
\PassOptionsToPackage{normalem}{ulem}
\usepackage{ulem}

\makeatletter

\providecolor{lyxadded}{rgb}{0,0,1}
\providecolor{lyxdeleted}{rgb}{1,0,0}

\@ifundefined{textcolor}{}
{%
 \definecolor{BLACK}{gray}{0}
 \definecolor{WHITE}{gray}{1}
 \definecolor{RED}{rgb}{1,0,0}
 \definecolor{GREEN}{rgb}{0,1,0}
 \definecolor{BLUE}{rgb}{0,0,1}
 \definecolor{CYAN}{cmyk}{1,0,0,0}
 \definecolor{MAGENTA}{cmyk}{0,1,0,0}
 \definecolor{YELLOW}{cmyk}{0,0,1,0}
}

\usepackage{babel}

\makeatother

\usepackage{babel}
\begin{document}
\global\long\def\oc#1{\hat{c}_{#1}}
 \global\long\def\ocd#1{\hat{c}_{#1}^{\dagger}}
 \global\long\def\tr{\text{Tr}\,}
 \global\long\def\im{\text{Im}\,}
 \global\long\def\re{\text{Re}\,}
 \global\long\def\bra#1{\left\langle #1\right|}
 \global\long\def\ket#1{\left|#1\right\rangle }
 \global\long\def\braket#1#2{\left.\left\langle #1\right|#2\right\rangle }
 \global\long\def\obracket#1#2#3{\left\langle #1\right|#2\left|#3\right\rangle }
 \global\long\def\proj#1#2{\left.\left.\left|#1\right\rangle \right\langle #2\right|}

\title{Dynamics of Many-Body Localization}

\author{Yevgeny Bar Lev}

\author{David R. Reichman}

\affiliation{Department of Chemistry, Columbia University, 3000 Broadway, New
York, New York 10027, USA }
\begin{abstract}
Following the field theoretic approach of Basko \emph{et al}., Ann.
Phys. \textbf{321}, 1126 (2006), we study in detail the real-time
dynamics of a system expected to exhibit many-body localization. In
particular, for time scales inaccessible to exact methods, we demonstrate
that within the second-Born approximation that the temporal decay
of the density-density correlation function is non-exponential and
is consistent with a finite value for $t\to\infty$, as expected in
a non-ergodic state. This behavior persists over a wide range of disorder
and interaction strengths. We discuss the implications of our findings
with respect to dynamical phase boundaries based both on exact diagonalization
studies and as well as those established by the methods of Ref.~\onlinecite{Basko2006a}.
\end{abstract}
\maketitle
It has been known for more than 50 years that \emph{non-interacting}
particles in a one-dimensional disordered system exhibit Anderson
localization \cite{Anderson1958b}, namely the exponential suppression
of transport. While a localized system is non-ergodic and thus does
not thermalize, coupling the system to other degrees of freedom with
a continuous spectrum, such as a heat bath, allows thermalization
to occur via processes such as variable-range hopping \cite{Mott1969}.
For an isolated many-body system, only interactions between the particles
may lead to thermalization. The question of whether or not localization
is stable in the presence of interactions was first considered by
Fleishman and Anderson \cite{Fleishman1980a}, who concluded that
short-ranged interactions cannot destabilize the insulating phase.
A similar and still open question also exists for Bose-Einstein condensates,
treated in the framework of the time-dependent Gross-Pitaevskii (or
nonlinear Schrödinger) equation \cite{Fishman2009b}. In this case
numerics, as well as analytical arguments, suggest a temporally sub-diffusive
or even logarithmic thermalization behavior for not very strong interactions
\cite{Fishman2009b}.

Using a diagrammatic approach, Basko \emph{et al.} argued that for
a general class of isolated, disordered and interacting systems, a
many-body mobility edge exists, similarly to the Anderson mobility
edge in a three-dimensional non-interacting system \cite{Basko2006a}.
Namely, a critical energy separates ``insulating'' and ``metallic''
eigenstates, which can be distinguished by evaluating the spatial
correlations of any local operator. ``Metallic'' eigenstates will
have non-vanishing or slowly decaying correlations, while ``insulating''
states will have exponentially decaying correlations. By changing
the energy (or the micro canonical temperature) of the system across
the mobility-edge, the system will undergo an insulator--metal transition.
Similar to the Anderson transition, the many-body localization (MBL)
transition is a dynamical and \emph{not} a thermodynamic phenomena
\cite{Basko2006a}. However, the MBL transition is also not a conventional
quantum phase transition since the critical energy, which depends
on the parameters of the system, may be very far from the ground state.
In fact, for systems of bounded energy density (e.g., finite number
of states per site), Oganesyan and Huse suggested that the transition
will persist up to infinite temperature \cite{Oganesyan2007a}. Namely,
nontrivial parameters of the system may be found such that essentially
\emph{all }the eigenstates are ``insulating''. For a zero dimensional
system mapped to the Bethe lattice, it was theoretically proposed
\cite{Altshuler1997} and recently numerically examined \cite{DeLuca2013},
that for some range of parameters the metallic phase can be \emph{non-ergodic}.
The existence of a non-ergodic \emph{metallic} phase for \emph{finite}
dimensional systems had been conjectured \cite{Altshuler2010}, but
has been numerically tested only for small systems \cite{Luca2013}.

By calculation of the DC conductivity or the properties of eigenfunctions
for sufficiently small systems, the MBL transition has gained support
from numerical studies that utilize either direct diagonalization
\cite{Oganesyan2007a,Karahalios2009a,Pal2010a,Berkelbach2010a,Barisic2010a}
or methods of similar numerical complexity \cite{Monthus2010a}. However,
all of these studies suffer from the drawback that the numerically
accessible system size is about $16$ sites, which does not allow
for a systematic analysis of finite size effects. This is not an issue
deep within the insulating phase, but due to the divergence of the
interacting localization length at the transition, it introduces severe
difficulties for examining the system near the putative transition
and in the metallic phase. Other studies have examined the dynamical
nature of the transition using time-dependent density matrix renormalization
group (tDMRG) or similar methods \cite{Znidaric2008,Bardarson2012}.
However, these studies are restricted to the localized phase due to
the growth of entanglement entropy. While the existence of an insulating
phase for some range of parameters and energies appears to be quite
well established even at the rigorous level \cite{Aizenman2009b},
this is not the case for the metallic phase, where there are currently
no rigorous results and the existing numerical schemes are quite limited.

In this letter, we examine the dynamics of an isolated one-dimensional
system across the putative MBL transition predicted by exact diagonalization
studies of the same system. In particular, we study the relaxation
dynamics of our system starting from a far-from-equilibrium initial
condition that is a pure state of the corresponding non-interacting
system. We follow the diagrammatic approach of Ref. \onlinecite{Basko2006a},
while relaxing several assumptions used in that work. Unlike the work
of Ref. \onlinecite{Basko2006a}, we compute in detail the dynamics
of the system from an appropriately chosen initial condition. By doing
so, we are able to assess the accuracy of the approximations used
in Ref.~\onlinecite{Basko2006a} against exact numerical results
(where available) as well as to describe how the MBL transition should
manifest within the very framework that first predicted its existence. 

\begin{figure}
\begin{centering}
\includegraphics[clip,width=8.6cm]{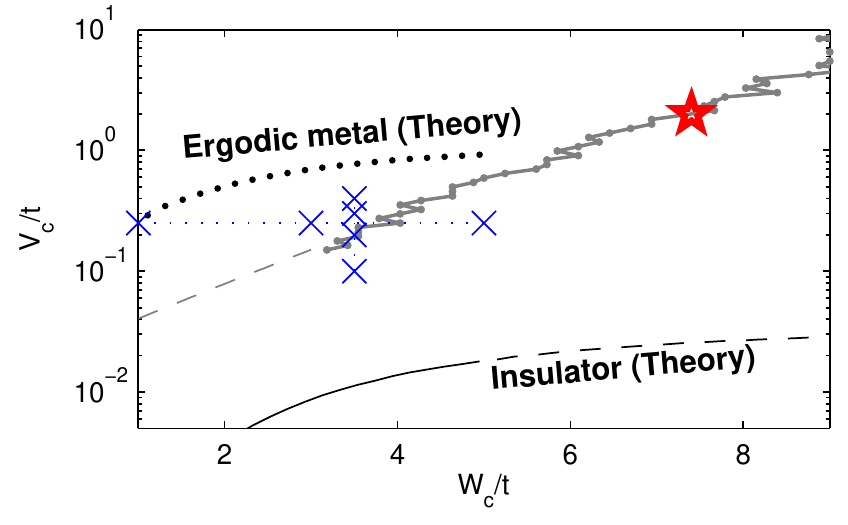} 
\par\end{centering}

\caption{\label{fig:params}(color online) Dynamical phase boundaries of the
system as a function of $W/t$ and $V/t$. Gray solid line indicates
the numerical critical values calculated using the method of Ref.~\onlinecite{Pal2010a}
($N=12$, see their Fig. 6). The red star represents the transition
point obtained in Ref.~\onlinecite{Pal2010a} (note the factor of
2 difference in $W_{c}$). The blue crosses represent the parameters
used in this work. Vertical cut: $W=3.5$ and $V=0.1,\,0.2\,0.3,\,0.4$.
Horizontal cut: $W=1,3,5$ and $V=0.25$. The solid black line is
the solution of Eq.~\eqref{eq:BAA}, and separates metal from insulator.
The dotted black line is obtained from Eq.~\eqref{eq:BAA0}, and
demarcates the boundary between ergodic and non-ergodic metals. The
dashed lines are extrapolations to regions below which such equations
are \emph{not} valid.}
\end{figure}

Following previous studies \cite{Znidaric2008,Karahalios2009a,Barisic2010a,Berkelbach2010a,Monthus2010a,Pal2010a},
we investigate a one-dimensional system of spinless and interacting
fermions in a disordered potential,

\begin{eqnarray}
H & = & -t\sum_{i}\left(\hat{c}_{i}^{\dagger}\hat{c}_{i+1}+\hat{c}_{i+1}^{\dagger}\hat{c}_{i}\right)\label{eq:t-V_model}\\
 & + & V\sum_{i}\left(\hat{n}_{i}-\frac{1}{2}\right)\left(\hat{n}_{i+1}-\frac{1}{2}\right)+\sum_{i}h_{i}\left(\hat{n}_{i}-\frac{1}{2}\right),\nonumber 
\end{eqnarray}
where $t$ is the hopping matrix element, $V$ is the interaction
strength and $h_{i}$ are random fields independently distributed
on the interval $h_{i}\in\left[-W,W\right]$. Note that by using the
Jordan-Wigner transformation, this model can be exactly mapped onto
the XXZ model. There are only two independent parameters in the Hamiltonian
and we therefore choose $t=1$. Since this model has a bounded energy
density $\left|\left\langle H\right\rangle \right|/L\leq\left(2t+V/4+W/2\right)$
(where $L$ is the length of the system), critical parameters may
be found such that the system will transition from a mixture of ``insulating''
and ``metallic'' states to a situation where \emph{all }of the many-body
eigenstates are ``insulating.'' The existence of such parameters
is equivalent to the assumption used in Ref.~\onlinecite{Oganesyan2007a}
that the MBL transition will survive at infinite temperature. Under
these conditions, the critical disorder strength has been determined
in previous studies for rather strong interaction $V=2t$, to be about
$W_{c}/t\approx7-8$ \cite{Berkelbach2010a,Pal2010a}. Since our approach
is based on many-body perturbation theory, we are limited to small
interaction strengths. Therefore, to find the relevant critical parameters,
we extend the calculations of Ref.~\onlinecite{Pal2010a} to evaluate
the critical \emph{line} in the space of $W/t$ and $V/t$ using exact
diagonalization (see Fig. \ref{fig:params}). We also estimate the
theoretical dynamical phase boundaries by using the critical temperatures
and a thermodynamic relation \cite{Basko2006a,Aleiner2013} 
\begin{equation}
T_{el}=\frac{t^{2}}{\nu\xi V^{2}},\qquad T_{c}=\frac{t}{12\nu\xi V\ln t/V},\qquad\frac{E}{L}=\frac{\pi^{2}}{6}\nu T^{2}.\label{eq:BAA0}
\end{equation}
Here $\xi$ is the non-interacting localization length, $T_{el}$
is the temperature which separates ergodic and non-ergodic metals
(see e.g., \cite{Altshuler2010}), $T_{c}$ is the transition temperature
and $\nu$ is one-particle density of states. Although for high temperatures
and for a system with a bounded energy density the thermodynamic relation
is not strictly valid, we use it to extrapolate from low to high temperatures
by setting $E_{c}/L$ to be equal to half of the energy band $\Delta\thickapprox\left(2t+W/2\right)$
(the interaction contribution is negligible) \cite{Aleiner2013}.
The critical interaction can now be obtained in terms of the non-interacting
localization length, $\xi\left(W\right)$ and $W$, \textbf{
\begin{equation}
\frac{V_{el}}{t}=\frac{1}{\sqrt{\alpha}},\qquad\frac{V_{c}}{t}=\frac{1}{12\alpha\left(\ln12\alpha\left(\ln\left(12\alpha\cdots\right)\right)\right)},\label{eq:BAA}
\end{equation}
}where $\alpha=\sqrt{6\nu\Delta\xi^{2}/\left(\pi^{2}t\right)}$. The
theoretical lines are valid for $V_{c}/t<1$ and $\xi>1,$ and are
plotted at Fig. \ref{fig:params}. It is clearly seen that the numerical
critical line is one order of magnitude higher than the theoretical
one, lying predominately in the non-ergodic metal phase. We will show,
however, that at least for small interaction strengths the numerical
critical line suffers from severe finite size effects. Taking this
into account suggests that the line should move towards higher values
of $V$, which will drive it even further away from the phase boundary
where theory predicts a stable insulator. One possible explanation
of this discrepancy could be that the the insulating phase is stable
up to $T_{el}$ and not $T_{c}$ as predicted by the theory. In other
words, the non-ergodic metal phase is also insulating. Alternatively,
the approach used to determine the numerical critical line \cite{Pal2010a}
might be sensitive to an ergodic non-ergodic transition within the
metal phase and not the true metal-insulator transition.

We next outline our dynamic scheme which is similar in spirit to Ref.~\onlinecite{Basko2006a},
but relaxes several approximations of that work. We start with the
one-particle greater and lesser non-equilibrium Green's functions,
\begin{eqnarray}
G_{ij}^{>}\left(t;t'\right) & = & -i\tr\left\{ \hat{\rho}_{0}\hat{c}_{i}\left(t\right)\hat{c}_{j}^{\dagger}\left(t'\right)\right\} \label{eq:gtr_less_G}\\
G_{ij}^{<}\left(t;t'\right) & = & i\tr\left\{ \hat{\rho}_{0}\hat{c}_{j}^{\dagger}\left(t'\right)\hat{c}_{i}\left(t\right)\right\} ,\nonumber 
\end{eqnarray}
where $\hat{\rho}_{0}$ is the initial density matrix. For a non-interacting
initial density matrix, the Green's functions obey the Kadanoff--Baym
equations of motion \cite{Kadanoff1994}, 
\begin{eqnarray}
i\partial_{t}G^{\gtrless}\left(t,t'\right) & = & \left(\hat{h}_{0}+\Sigma^{HF}\left(t\right)\right)G^{\gtrless}\left(t,t'\right)\nonumber \\
 & + & \int_{0}^{t}\Sigma^{R}\left(t,t_{2}\right)G^{\gtrless}\left(t_{2},t'\right)\mathrm{d}t_{2}\nonumber \\
 & + & \int_{0}^{t'}\Sigma^{\gtrless}\left(t,t_{2}\right)G^{A}\left(t_{2},t'\right)\mathrm{d}t_{2},\label{eq:KB_eq}
\end{eqnarray}
where spatial indices and summations are suppressed for clarity; $\hat{h}_{0,nm}=-t\left(\delta_{n,m+1}+\delta_{n,m-1}\right)+h_{n}\delta_{nm}$
is the one particle Hamiltonian; $\Sigma^{HF}\left(t\right)$, $\Sigma^{\gtrless}\left(t\right)$
are the Hartree-Fock greater and lesser self-energies of the problem
respectively; and the superscripts 'R' and 'A' represent retarded
and advanced Green's functions and self-energies, which are defined
as 
\begin{eqnarray}
\Sigma^{R}\left(t,t_{2}\right) & = & \theta\left(t-t_{2}\right)\left(\Sigma^{>}\left(t,t_{2}\right)-\Sigma^{<}\left(t,t_{2}\right)\right)\\
G^{A}\left(t_{2},t'\right) & = & -\theta\left(t'-t_{2}\right)\left(G^{>}\left(t_{2},t'\right)-G^{<}\left(t_{2},t'\right)\right).\nonumber 
\end{eqnarray}
Note that due to the complexity of (\ref{eq:KB_eq}), what was actually
considered in Ref.~\onlinecite{Basko2006a} is the corresponding
quantum Boltzmann equation. For this purpose the authors have neglected
the off-diagonal spatial elements of the Green's functions, and performed
a gradient expansion of the time variable. Additionally, the real
part of the self-energy as well as the Hartree-Fock contributions
have been neglected \cite{Basko2006a}. Although it is numerically
feasible to solve (\ref{eq:KB_eq}) within the specified approximation
for the self-energies (see below), this approach is turns unstable
for sufficiently long times of any parameters of the Hamiltonian.
To eliminate this spurious behavior we reduce (\ref{eq:KB_eq}) to
a quantum master equation for the one-particle density matrix by introduction
of the generalized Kadanoff-Baym anzatz, 
\begin{equation}
G^{\lessgtr}\left(t,t'\right)=i\left[G^{R}\left(t,t'\right)G^{\lessgtr}\left(t',t'\right)-G^{\lessgtr}\left(t,t\right)G^{A}\left(t,t'\right)\right],\label{eq:GKBA}
\end{equation}
and by approximating the retarded and advanced Green's functions with
their Hartree-Fock (HF) values \cite{Spicka2005,Spicka2005a,Latini2013}.
This approach is in the spirit of the Boltzmann approach of Ref.~\onlinecite{Basko2006a},
and is ostensibly more precise, since it considers the full density
matrix and not just its diagonal values. Additionally, the quasi-classical
approximation as well as gradient expansions are not needed, which
allows us to describe systems far from equilibrium. It is not our
purpose to examine the validity of this approximation in this letter
and the reader is referred to Refs.~\onlinecite{Spicka2005,Spicka2005a}.
We leave the discussion of the full solution of (\ref{eq:KB_eq})
to a future study. As in Ref.~\onlinecite{Basko2006a} we utilize
both the HF and the self-consistent second-Born (2B, called SCBA in
\cite{Basko2006a}) approximations for the self energies, 
\begin{eqnarray}
\Sigma_{ij}^{HF}\left(t\right) & = & -i\delta_{ij}\sum_{k}V_{ik}G_{kk}^{<}\left(t;t\right)+iV_{ij}G_{ij}^{<}\left(t;t\right)\nonumber \\
\Sigma_{ij}^{>}\left(t,t'\right) & = & \sum_{k,l}V_{il}V_{jk}G_{kl}^{<}\left(t',t\right)\times\label{eq:self-energies}\\
 &  & \left[G_{lk}^{>}\left(t,t'\right)G_{ij}^{>}\left(t,t'\right)-G_{lj}^{>}\left(t,t'\right)G_{ik}^{>}\left(t,t'\right)\right],\nonumber 
\end{eqnarray}
where $V_{ij}=V\left(\delta_{i,j+1}+\delta_{i,j-1}\right)$ is the
interaction potential.

The fact that the insulating phase is non-ergodic means that great
care must be exercised with regards to the choice of the initial conditions
\cite{Aleiner2013,Basko2006a}. To illustrate the issues involved,
consider a disconnected interacting system, namely by setting $t=0$.
This system can be solved exactly, given the fact that the Hamiltonian
is already diagonal in the position basis. A similar model was studied
by several authors in the context of the MBL, but for distinct purposes,
see Refs.~\onlinecite{Huse2013,Serbyn2013}. The Green's function
is a periodic function of time with period $V$. Ignoring the fact
that the system is solvable and utilizing the diagrammatic perturbation
theory up to second order in $V$ (as in the 2B approximation) gives,
\begin{eqnarray}
\Sigma_{i}^{<}\left(t\right) & = & i\left\langle \hat{n}_{i}\right\rangle e^{-i\varepsilon_{i}t}\sum_{j}V_{i,j}\left\langle \hat{n}_{j}\right\rangle \left\langle 1-\hat{n}_{j}\right\rangle ,\label{eq:self-energy-disconn}
\end{eqnarray}
where $\left\langle \hat{n}_{j}\right\rangle =\tr\hat{\rho}_{0}\hat{n}_{j}$.
Notably, the imaginary part of the self-energy yields an unphysical
decay of the Green's function, and only vanishes if $\left\langle \hat{n}_{i}\right\rangle =n_{i}=0,1$
(which implies a density matrix which is one Slater determinant localized
on lattice sites). It can be shown that with this form of initial
density matrix the problem may be solved using only the Hartree term.
Thus, if we would like to recover the proper $t\to0$ limit in the
framework of 2B, any selected initial state should have the properties
disclosed above in this limit. To satisfy this restriction, we use
the following initial condition: 
\begin{equation}
G_{ij}^{<}\left(0,0\right)=i\sum_{k}\phi_{k}\left(i\right)\phi_{k}\left(j\right)n_{k}^{0},\label{eq:init_cond}
\end{equation}
where $n_{k}^{0}\in\left(0,1\right)$, and $\phi_{k}\left(i\right)$
are the non-interacting one-particle eigenstates. We consider half-filling
throughout this letter, although other fillings have been investigated
and do not lead to distinct behavior.

For a MBL transition at finite temperature, the mobility edge is found
within the many-body spectrum, and therefore the average energy as
well as the energetic width of the initial condition are of great
importance, since they determine the position of the state with respect
to the many-body mobility edge. For a MBL transition at infinite temperature,
for certain parameters, \emph{all }eigenstates become ``insulating,''
and therefore the initial condition should not be important. Nevertheless,
to eliminate one control parameter, we set the location of the mean
energy density of the initial condition to be in the middle of the
many-body energy band, namely $\left\langle H\right\rangle =0$. This
closely corresponds to the $T\to\infty$ limit invoked in Ref.~\onlinecite{Oganesyan2007a},
and allows one to establish the onset of the transition when the mobility
edge converges to zero. The relaxation of the system is monitored
by calculating the correlation function 
\begin{equation}
\delta\rho\left(t\right)=\frac{1}{L}\sum_{k}\overline{\left\langle \left(\hat{n}_{k}\left(t\right)-\frac{1}{2}\right)\left(\hat{n}_{k}\left(0\right)-\frac{1}{2}\right)\right\rangle },\label{eq:measure}
\end{equation}
where $\hat{n}_{k}=\sum_{ij}\phi_{k}\left(i\right)\phi_{k}\left(j\right)\hat{c}_{i}^{\dagger}\hat{c}_{j}$,
and the over-line indicates averaging over disorder realizations.
Generically a two-particle Green's function is needed to calculate
this quantity, however the chosen initial condition \eqref{eq:init_cond}
renders $\left\langle \hat{n}_{k}\left(t\right)\hat{n}_{k}\left(0\right)\right\rangle =\left\langle \hat{n}_{k}\left(t\right)\right\rangle \left\langle \hat{n}_{k}\left(0\right)\right\rangle $.
Since the total number of particles is conserved, this correlation
function measures the diffusion rate of the one-particle energy, and
for an initial conditions in the ergodic phase it will typically decay
as $t^{-1/2}$ or faster as a function of time (see, e.g., Ref.~\onlinecite{Fabricius1998}
for studies of clean systems). For initial conditions in a non-ergodic
phase temporal decay will cease after some finite time, or alternatively,
a sub-diffusive relaxation is expected. Note that although a non-decaying
correlation indicates that some of the particles are pinned to their
initial positions, it does not preclude a finite mobility for the
rest of the particles, and therefore does not rule out metallic behavior.
Nevertheless, in this case the mobility of the particles is expected
to be considerably impaired with comparison to the ergodic case. It
is precisely this non-ergodic conducting situation which defines the
non-ergodic metal region of the dynamic phase diagram of Fig. \ref{fig:params}.

\begin{figure}
\begin{centering}
\includegraphics[width=8.6cm]{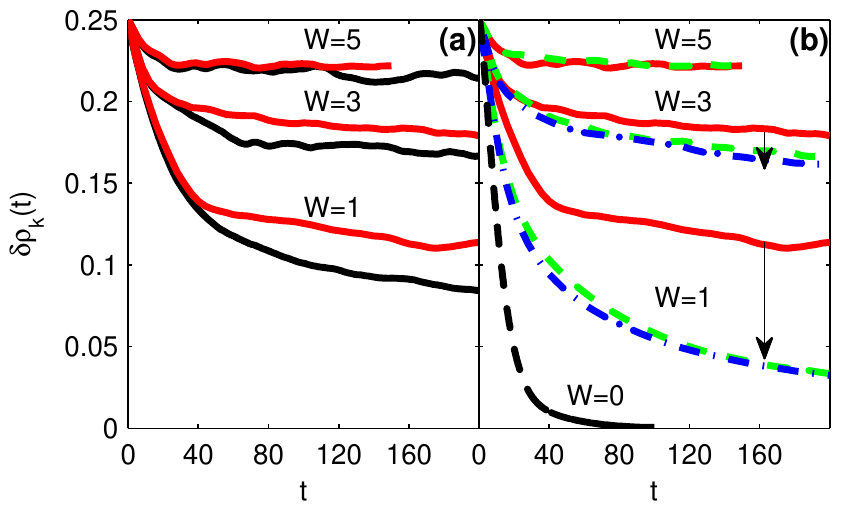} 
\par\end{centering}

\caption{\label{fig:exact-2B}(color online) \textbf{(a)} Density-density correlation
function as a function of time for averaged over $256$ disorder realizations
$\left(V=0.25\right)$. The solid black lines designate the exact
solution calculated using ED $\left(L=12\right)$ and red (gray) lines
designate the solution using QME. \textbf{(b)} Finite size behavior
for same parameters using QME. The system sizes used: $W=0,$ $N=2^{14}$
(black dashed); $W=1,$ $N=12,48,96$ (solid, dashed, dot dashed);
$W=3,$ $N=12,28,36$ (solid, dashed, dot dashed) and $W=5,$ $N=12,28$
(solid, dashed).}
\end{figure}

We solve the quantum master equation (QME) numerically for the one-particle
Green's functions (\ref{eq:KB_eq}) as a function of time, with the
initial conditions (\ref{eq:init_cond}) and the self-energies (\ref{eq:self-energies}).
For this purpose we use the numerical method developed in \cite{Stan2009}.
The correlation function is averaged over 256 realizations of the
disordered potential. In Fig. \ref{fig:exact-2B}a we compare the
perturbative calculation to the exact solution of a small chain, $L=12$,
obtained by exact diagonalization (ED). A remarkable correspondence
between the QME and the exact solutions is seen for $V=0.25$ and
times $t\lesssim40$. This correspondence becomes better for larger
$W$ and does not exist at the HF level, which produces only non-decaying
solutions (not shown). For longer times there is only a qualitative
correspondence between the QME and the exact solution. It should also
be noted that for smaller values of $W$ such as $W=1$ (where the
non-interacting localization length, $\xi=25$, is larger than the
system size) the $L=12$ system is simply a model for testing the
approximation scheme and has little bearing on the dynamics of MBL
in the thermodynamic limit, although surprisingly the dynamics is
not very different from that at much larger disorder strengths.

\begin{figure}
\centering{}\includegraphics[clip]{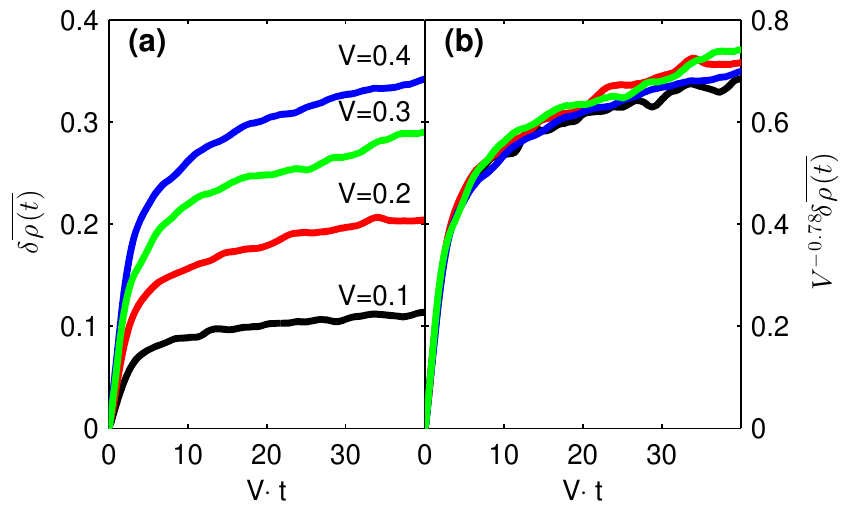} \caption{\label{fig:rescaling}(color online) \textbf{(a)} Modified density-density
correlation $\left(\delta\bar{\rho}\left(t\right)\equiv1-4\delta\rho\left(t\right)\right)$
as a function of $V\cdot t$ , averaged over $256$ disorder realizations,
for $W=3.5,\, L=36$. \textbf{(b)} Left panel rescaled by plotting
$V^{-0.78}\delta\bar{\rho}$ as a function of $V\cdot t$ (see text).}
\end{figure}
From Fig.~\ref{fig:exact-2B}b we observe that after the initial
relaxation there is a slow decay of the correlation function. The
decay of the clean system is exponential with a time scale of $t_{1}\sim V^{-2}$,
which indicates that the QME within the 2B approximation \emph{overestimates}
the relaxation in the clean (and presumably also in the nearly metallic)
region of the phase diagram \cite{Fabricius1998}. To eliminate finite
size effects we have increased the size of the system until no changes
are observed in our measure. Note that, up to the considered time,
$L=12$ suffices only for $W>3$, which suggests that the numerically
obtained critical line of Fig. \ref{fig:params} is prone to large
finite size effects, at least for the considered interaction strengths.
Thus, if that line exists in the $L\to\infty$ limit, it should move
toward higher values of $V$ in the $V-W$ plane.

Although at the MBL transition we expect to observe a steep change
in the functional dependence of the correlation functions, the behavior
of density fluctuations appear surprisingly smooth over a broad range
of parameters. To demonstrate this, we fix $W$ and cross the numerically
determined critical line by changing $V$ (see Fig. \ref{fig:params}).
The correlation function can be reasonably well rescaled by setting
$\delta\bar{\rho}\left(t\right)\equiv1-4\delta\rho\left(t\right)$,
while plotting $V^{-0.78}\delta\bar{\rho}\left(t\right)$ as a function
of rescaled time, $V\cdot t$ (see Fig. \ref{fig:rescaling}). The
exponent was obtained by fitting. We could not find satisfactory rescaling
for fixed $V$. The initial fast relaxation time may be inferred from
Fig. \ref{fig:rescaling}, $t_{1}\sim6V^{-1}$. During this time the
system dephases across the exact many-body states, which span the
initial state. The scaling suggests that, $\delta\rho\left(t\right)=\frac{1}{4}-AV^{0.78}f\left(Vt\right)$,
and therefore assuming its asymptotic validity it should decay to
zero at the ergodic-non-ergodic transition $V=V_{*}$. This yields
the form $\delta\rho\left(V,t\to\infty\right)=\frac{1}{4}\left(1-\left(V/V_{*}\right)^{0.78}\right)$.
In particular, the smooth character of the temporal behavior suggests
that $V_{*}>0.4$. \emph{Thus our numerics suggest that all correlation
functions in the region of investigation asymptotically decay to a
finite value corresponding to a non-ergodic state.}

To summarize, this work presents, for the first time, a detailed description
of the dynamical phase diagram and of how dynamical quantities manifest
in the MBL scenario. Although our approach is approximate, it is based
on the same approximations that first predicted the existence of MBL
\cite{Basko2006a}, and thus should provide important qualitative
guidelines to the long-time behavior that is out of reach by more
rigorous methods. We find that for all values of interaction and disorder
strengths studied, density fluctuations decay in a remarkably slow,
non-exponential manner. Furthermore, the dynamics do not qualitatively
change in a broad region of parameter space, and are consistent with
a non-ergodic phase. This is rather surprising, given that our scan
of parameters takes the system across the transition line of Ref.~\onlinecite{Pal2010a}
and in the vicinity of the ergodic metal region proposed in Ref.~
\onlinecite{Basko2006a}. Our results, however, may be viewed as consistent
with both works, if the phase boundary of Ref.~\onlinecite{Pal2010a},
 which based on our finite size analysis is expected to drift upward
in the thermodynamic limit, demarcates an ``ergodic-non-ergodic''
transition where the non-ergodic phase is actually a non-ergodic metal.
To resolve these questions would require calculation of the conductivity
in addition to the density fluctuations. This is much more difficult
to do within the approach described here. Research along these lines
will be presented in a future publication.

We would like to thank I. Aleiner, O. Agam, D. Huse and M. Schiro
for many enlightening and helpful discussions. This work used the
Extreme Science and Engineering Discovery Environment (XSEDE), which
is supported by National Science Foundation grant number OCI-1053575.
This work was supported by the Fulbright foundation and by grant NSF-CHE-1213247.

\bibliographystyle{apsrev}
\bibliography{MBL}

\begin{thebibliography}{27}
\expandafter\ifx\csname natexlab\endcsname\relax\def\natexlab#1{#1}\fi
\expandafter\ifx\csname bibnamefont\endcsname\relax
  \def\bibnamefont#1{#1}\fi
\expandafter\ifx\csname bibfnamefont\endcsname\relax
  \def\bibfnamefont#1{#1}\fi
\expandafter\ifx\csname citenamefont\endcsname\relax
  \def\citenamefont#1{#1}\fi
\expandafter\ifx\csname url\endcsname\relax
  \def\url#1{\texttt{#1}}\fi
\expandafter\ifx\csname urlprefix\endcsname\relax\def\urlprefix{URL }\fi
\providecommand{\bibinfo}[2]{#2}
\providecommand{\eprint}[2][]{\url{#2}}

\bibitem[{\citenamefont{Basko et~al.}(2006)\citenamefont{Basko, Aleiner, and
  Altshuler}}]{Basko2006a}
\bibinfo{author}{\bibfnamefont{D.}~\bibnamefont{Basko}},
  \bibinfo{author}{\bibfnamefont{I.~L.} \bibnamefont{Aleiner}},
  \bibnamefont{and}
  \bibinfo{author}{\bibfnamefont{B.}~\bibnamefont{Altshuler}},
  \bibinfo{journal}{Ann. Phys. (N. Y).} \textbf{\bibinfo{volume}{321}},
  \bibinfo{pages}{1126} (\bibinfo{year}{2006}).

\bibitem[{\citenamefont{Anderson}(1958)}]{Anderson1958b}
\bibinfo{author}{\bibfnamefont{P.~W.} \bibnamefont{Anderson}},
  \bibinfo{journal}{Phys. Rev.} \textbf{\bibinfo{volume}{109}},
  \bibinfo{pages}{1492} (\bibinfo{year}{1958}).

\bibitem[{\citenamefont{Mott}(1969)}]{Mott1969}
\bibinfo{author}{\bibfnamefont{N.~F.} \bibnamefont{Mott}},
  \bibinfo{journal}{Phil. Mag.} \textbf{\bibinfo{volume}{19}},
  \bibinfo{pages}{835} (\bibinfo{year}{1969}).

\bibitem[{\citenamefont{Fleishman and Anderson}(1980)}]{Fleishman1980a}
\bibinfo{author}{\bibfnamefont{L.}~\bibnamefont{Fleishman}} \bibnamefont{and}
  \bibinfo{author}{\bibfnamefont{P.~W.} \bibnamefont{Anderson}},
  \bibinfo{journal}{Phys. Rev. B} \textbf{\bibinfo{volume}{21}},
  \bibinfo{pages}{2366} (\bibinfo{year}{1980}).

\bibitem[{\citenamefont{Fishman et~al.}(2009)\citenamefont{Fishman, Krivolapov,
  and Soffer}}]{Fishman2009b}
\bibinfo{author}{\bibfnamefont{S.}~\bibnamefont{Fishman}},
  \bibinfo{author}{\bibfnamefont{Y.}~\bibnamefont{Krivolapov}},
  \bibnamefont{and} \bibinfo{author}{\bibfnamefont{A.}~\bibnamefont{Soffer}},
  \bibinfo{journal}{Nonlinearity} \textbf{\bibinfo{volume}{22}},
  \bibinfo{pages}{2861} (\bibinfo{year}{2009}).

\bibitem[{\citenamefont{Oganesyan and Huse}(2007)}]{Oganesyan2007a}
\bibinfo{author}{\bibfnamefont{V.}~\bibnamefont{Oganesyan}} \bibnamefont{and}
  \bibinfo{author}{\bibfnamefont{D.~A.} \bibnamefont{Huse}},
  \bibinfo{journal}{Phys. Rev. B} \textbf{\bibinfo{volume}{75}},
  \bibinfo{pages}{155111} (\bibinfo{year}{2007}).

\bibitem[{\citenamefont{Altshuler et~al.}(1997)\citenamefont{Altshuler, Gefen,
  Kamenev, and Levitov}}]{Altshuler1997}
\bibinfo{author}{\bibfnamefont{B.}~\bibnamefont{Altshuler}},
  \bibinfo{author}{\bibfnamefont{Y.}~\bibnamefont{Gefen}},
  \bibinfo{author}{\bibfnamefont{A.}~\bibnamefont{Kamenev}}, \bibnamefont{and}
  \bibinfo{author}{\bibfnamefont{L.}~\bibnamefont{Levitov}},
  \bibinfo{journal}{Phys. Rev. Lett.} \textbf{\bibinfo{volume}{78}},
  \bibinfo{pages}{2803} (\bibinfo{year}{1997}).

\bibitem[{\citenamefont{{De Luca} et~al.}(2013)\citenamefont{{De Luca},
  Scardicchio, Kravtsov, and Altshuler}}]{DeLuca2013}
\bibinfo{author}{\bibfnamefont{A.}~\bibnamefont{{De Luca}}},
  \bibinfo{author}{\bibfnamefont{A.}~\bibnamefont{Scardicchio}},
  \bibinfo{author}{\bibfnamefont{V.~E.} \bibnamefont{Kravtsov}},
  \bibnamefont{and} \bibinfo{author}{\bibfnamefont{B.~L.}
  \bibnamefont{Altshuler}}, p.~\bibinfo{pages}{8} (\bibinfo{year}{2013}),
  \eprint{1401.0019}, \urlprefix\url{http://arxiv.org/abs/1401.0019}.

\bibitem[{\citenamefont{Altshuler}(2010)}]{Altshuler2010}
\bibinfo{author}{\bibfnamefont{B.}~\bibnamefont{Altshuler}}
  (\bibinfo{year}{2010}),
  \urlprefix\url{http://www.lancaster.ac.uk/users/esqn/windsor10/lectures/Alts%
huler.pdf}.

\bibitem[{\citenamefont{Luca and Scardicchio}(2013)}]{Luca2013}
\bibinfo{author}{\bibfnamefont{A.~D.} \bibnamefont{Luca}} \bibnamefont{and}
  \bibinfo{author}{\bibfnamefont{A.}~\bibnamefont{Scardicchio}},
  \bibinfo{journal}{EPL} \textbf{\bibinfo{volume}{101}}, \bibinfo{pages}{37003}
  (\bibinfo{year}{2013}).

\bibitem[{\citenamefont{Karahalios et~al.}(2009)\citenamefont{Karahalios,
  Metavitsiadis, Zotos, Gorczyca, and Prelov\v{s}ek}}]{Karahalios2009a}
\bibinfo{author}{\bibfnamefont{A.}~\bibnamefont{Karahalios}},
  \bibinfo{author}{\bibfnamefont{A.}~\bibnamefont{Metavitsiadis}},
  \bibinfo{author}{\bibfnamefont{X.}~\bibnamefont{Zotos}},
  \bibinfo{author}{\bibfnamefont{A.}~\bibnamefont{Gorczyca}}, \bibnamefont{and}
  \bibinfo{author}{\bibfnamefont{P.}~\bibnamefont{Prelov\v{s}ek}},
  \bibinfo{journal}{Phys. Rev. B} \textbf{\bibinfo{volume}{79}},
  \bibinfo{pages}{024425} (\bibinfo{year}{2009}).

\bibitem[{\citenamefont{Pal and Huse}(2010)}]{Pal2010a}
\bibinfo{author}{\bibfnamefont{A.}~\bibnamefont{Pal}} \bibnamefont{and}
  \bibinfo{author}{\bibfnamefont{D.~A.} \bibnamefont{Huse}},
  \bibinfo{journal}{Phys. Rev. B} \textbf{\bibinfo{volume}{82}},
  \bibinfo{pages}{174411} (\bibinfo{year}{2010}).

\bibitem[{\citenamefont{Berkelbach and Reichman}(2010)}]{Berkelbach2010a}
\bibinfo{author}{\bibfnamefont{T.~C.} \bibnamefont{Berkelbach}}
  \bibnamefont{and} \bibinfo{author}{\bibfnamefont{D.~R.}
  \bibnamefont{Reichman}}, \bibinfo{journal}{Phys. Rev. B}
  \textbf{\bibinfo{volume}{81}}, \bibinfo{pages}{224429}
  (\bibinfo{year}{2010}).

\bibitem[{\citenamefont{Bari\v{s}i\'{c} and
  Prelov\v{s}ek}(2010)}]{Barisic2010a}
\bibinfo{author}{\bibfnamefont{O.~S.} \bibnamefont{Bari\v{s}i\'{c}}}
  \bibnamefont{and}
  \bibinfo{author}{\bibfnamefont{P.}~\bibnamefont{Prelov\v{s}ek}},
  \bibinfo{journal}{Phys. Rev. B} \textbf{\bibinfo{volume}{82}},
  \bibinfo{pages}{161106} (\bibinfo{year}{2010}).

\bibitem[{\citenamefont{Monthus and Garel}(2010)}]{Monthus2010a}
\bibinfo{author}{\bibfnamefont{C.}~\bibnamefont{Monthus}} \bibnamefont{and}
  \bibinfo{author}{\bibfnamefont{T.}~\bibnamefont{Garel}},
  \bibinfo{journal}{Phys. Rev. B} \textbf{\bibinfo{volume}{81}},
  \bibinfo{pages}{134202} (\bibinfo{year}{2010}).

\bibitem[{\citenamefont{\v{Z}nidari\v{c}
  et~al.}(2008)\citenamefont{\v{Z}nidari\v{c}, Prosen, and
  Prelov\v{s}ek}}]{Znidaric2008}
\bibinfo{author}{\bibfnamefont{M.}~\bibnamefont{\v{Z}nidari\v{c}}},
  \bibinfo{author}{\bibfnamefont{T.}~\bibnamefont{Prosen}}, \bibnamefont{and}
  \bibinfo{author}{\bibfnamefont{P.}~\bibnamefont{Prelov\v{s}ek}},
  \bibinfo{journal}{Phys. Rev. B} \textbf{\bibinfo{volume}{77}},
  \bibinfo{pages}{064426} (\bibinfo{year}{2008}).

\bibitem[{\citenamefont{Bardarson et~al.}(2012)\citenamefont{Bardarson,
  Pollmann, and Moore}}]{Bardarson2012}
\bibinfo{author}{\bibfnamefont{J.~H.} \bibnamefont{Bardarson}},
  \bibinfo{author}{\bibfnamefont{F.}~\bibnamefont{Pollmann}}, \bibnamefont{and}
  \bibinfo{author}{\bibfnamefont{J.~E.} \bibnamefont{Moore}},
  \bibinfo{journal}{Phys. Rev. Lett.} \textbf{\bibinfo{volume}{109}},
  \bibinfo{pages}{17202} (\bibinfo{year}{2012}).

\bibitem[{\citenamefont{Aizenman and Warzel}(2009)}]{Aizenman2009b}
\bibinfo{author}{\bibfnamefont{M.}~\bibnamefont{Aizenman}} \bibnamefont{and}
  \bibinfo{author}{\bibfnamefont{S.}~\bibnamefont{Warzel}},
  \bibinfo{journal}{Commun. Math. Phys.} \textbf{\bibinfo{volume}{290}},
  \bibinfo{pages}{903} (\bibinfo{year}{2009}).

\bibitem[{\citenamefont{Aleiner}(2013)}]{Aleiner2013}
\bibinfo{author}{\bibfnamefont{I.~L.} \bibnamefont{Aleiner}},
  \emph{\bibinfo{title}{{Private communication,}}} (\bibinfo{year}{2013}).

\bibitem[{\citenamefont{Kadanoff and Baym}(1994)}]{Kadanoff1994}
\bibinfo{author}{\bibfnamefont{L.~P.} \bibnamefont{Kadanoff}} \bibnamefont{and}
  \bibinfo{author}{\bibfnamefont{G.}~\bibnamefont{Baym}},
  \emph{\bibinfo{title}{{Quantum Statistical Mechanics}}}
  (\bibinfo{publisher}{Westview Press}, \bibinfo{year}{1994}), ISBN
  \bibinfo{isbn}{020141046X}.

\bibitem[{\citenamefont{\v{S}pi\v{c}ka
  et~al.}(2005{\natexlab{a}})\citenamefont{\v{S}pi\v{c}ka, Velick\'{y}, and
  Kalvov\'{a}}}]{Spicka2005}
\bibinfo{author}{\bibfnamefont{V.}~\bibnamefont{\v{S}pi\v{c}ka}},
  \bibinfo{author}{\bibfnamefont{B.}~\bibnamefont{Velick\'{y}}},
  \bibnamefont{and}
  \bibinfo{author}{\bibfnamefont{A.}~\bibnamefont{Kalvov\'{a}}},
  \bibinfo{journal}{Phys. E Low-dimensional Syst. Nanostructures}
  \textbf{\bibinfo{volume}{29}}, \bibinfo{pages}{154}
  (\bibinfo{year}{2005}{\natexlab{a}}).

\bibitem[{\citenamefont{\v{S}pi\v{c}ka
  et~al.}(2005{\natexlab{b}})\citenamefont{\v{S}pi\v{c}ka, Velick\'{y}, and
  Kalvov\'{a}}}]{Spicka2005a}
\bibinfo{author}{\bibfnamefont{V.}~\bibnamefont{\v{S}pi\v{c}ka}},
  \bibinfo{author}{\bibfnamefont{B.}~\bibnamefont{Velick\'{y}}},
  \bibnamefont{and}
  \bibinfo{author}{\bibfnamefont{A.}~\bibnamefont{Kalvov\'{a}}},
  \bibinfo{journal}{Phys. E Low-dimensional Syst. Nanostructures}
  \textbf{\bibinfo{volume}{29}}, \bibinfo{pages}{196}
  (\bibinfo{year}{2005}{\natexlab{b}}).

\bibitem[{\citenamefont{Latini et~al.}(2013)\citenamefont{Latini, Perfetto,
  Uimonen, van Leeuwen, and Stefanucci}}]{Latini2013}
\bibinfo{author}{\bibfnamefont{S.}~\bibnamefont{Latini}},
  \bibinfo{author}{\bibfnamefont{E.}~\bibnamefont{Perfetto}},
  \bibinfo{author}{\bibfnamefont{A.~M.} \bibnamefont{Uimonen}},
  \bibinfo{author}{\bibfnamefont{R.}~\bibnamefont{van Leeuwen}},
  \bibnamefont{and}
  \bibinfo{author}{\bibfnamefont{G.}~\bibnamefont{Stefanucci}},
  p.~\bibinfo{pages}{13} (\bibinfo{year}{2013}), \eprint{1311.4691},
  \urlprefix\url{http://arxiv.org/abs/1311.4691}.

\bibitem[{\citenamefont{Huse and Oganesyan}(2013)}]{Huse2013}
\bibinfo{author}{\bibfnamefont{D.~A.} \bibnamefont{Huse}} \bibnamefont{and}
  \bibinfo{author}{\bibfnamefont{V.}~\bibnamefont{Oganesyan}},
  \bibinfo{journal}{arXiv:1305.4915}  (\bibinfo{year}{2013}).

\bibitem[{\citenamefont{Serbyn et~al.}(2013)\citenamefont{Serbyn, Papi\'{c},
  and Abanin}}]{Serbyn2013}
\bibinfo{author}{\bibfnamefont{M.}~\bibnamefont{Serbyn}},
  \bibinfo{author}{\bibfnamefont{Z.}~\bibnamefont{Papi\'{c}}},
  \bibnamefont{and} \bibinfo{author}{\bibfnamefont{D.~A.}
  \bibnamefont{Abanin}}, \bibinfo{journal}{arXiv:1305.5554}
  (\bibinfo{year}{2013}).

\bibitem[{\citenamefont{Fabricius}(1998)}]{Fabricius1998}
\bibinfo{author}{\bibfnamefont{K.}~\bibnamefont{Fabricius}},
  \bibinfo{journal}{Phys. Rev. B} \textbf{\bibinfo{volume}{57}},
  \bibinfo{pages}{8340} (\bibinfo{year}{1998}).

\bibitem[{\citenamefont{Stan et~al.}(2009)\citenamefont{Stan, Dahlen, and van
  Leeuwen}}]{Stan2009}
\bibinfo{author}{\bibfnamefont{A.}~\bibnamefont{Stan}},
  \bibinfo{author}{\bibfnamefont{N.~E.} \bibnamefont{Dahlen}},
  \bibnamefont{and} \bibinfo{author}{\bibfnamefont{R.}~\bibnamefont{van
  Leeuwen}}, \bibinfo{journal}{J. Chem. Phys.} \textbf{\bibinfo{volume}{130}},
  \bibinfo{pages}{224101} (\bibinfo{year}{2009}).

\end{thebibliography}

\end{document}